# Phonon-mediated room-temperature quantum Hall transport in graphene

Daniel Vaquero[1,†], Vito Clericò[1,†], Michael Schmitz[2,3], Juan Antonio Delgado-Notario[1,4], Adrian Martín-Ramos[1], Juan Salvador-Sánchez[1], Claudius S. A. Müller[5,6], Km Rubi[5,6], Kenji Watanabe[7], Takashi Taniguchi[8], Bernd Beschoten[2], Christoph Stampfer[2,3], Enrique Diez[1], Mikhail I. Katsnelson[6], Uli Zeitler[5,6], Steffen Wiedmann[5,6], Sergio Pezzini[9,*]

[1]Nanotechnology Group, USAL–Nanolab, Universidad de Salamanca, E-37008 Salamanca, Spain.
[2]JARA-FIT and 2nd Institute of Physics, RWTH Aachen University, 52074 Aachen, Germany.
[3]Peter Grünberg Institute (PGI-9), Forschungszentrum Jülich, 52425 Jülich, Germany.
[4]CENTERA Laboratories, Institute of High Pressure Physics, Polish Academy of Sciences, 29/37 Sokołowska Str, Warsaw, Poland.
[5]High Field Magnet Laboratory (HFML-EMFL), Radboud University, Toernooiveld 7, 6525 ED Nijmegen, The Netherlands.
[6]Radboud University, Institute for Molecules and Materials, Heyendaalseweg 135, 6525 AJ Nijmegen, The Netherlands.
[7]Research Center for Functional Materials, National Institute for Materials Science, 1-1 Namiki Tsukuba, Ibaraki 305-0044, Japan.
[8]International Center for Materials Nanoarchitectonics, National Institute for Materials Science, 1-1 Namiki Tsukuba, Ibaraki 305-0044, Japan.
[9]NEST, Istituto Nanoscienze-CNR and Scuola Normale Superiore, Piazza San Silvestro 12, 56127 Pisa, Italy.

[†]These authors contributed equally to this work
[*]email: sergio.pezzini@nano.cnr.it

**Abstract**

The quantum Hall (QH) effect in two-dimensional electron systems (2DESs) is conventionally observed at liquid-helium temperatures, where lattice vibrations are strongly suppressed and bulk carrier scattering is dominated by disorder. However, due to large Landau level (LL) separation (~2000 K at $B$ = 30 T), graphene can support the QH effect up to room temperature (RT), concomitant with a non-negligible population of acoustic phonons with a wave-vector commensurate to the inverse electronic magnetic length. Here, we demonstrate that graphene encapsulated in hexagonal boron nitride (hBN) realizes a novel transport regime, where dissipation in the QH phase is governed predominantly by electron-phonon scattering. Investigating thermally-activated transport at filling factor 2 up to RT in an ensemble of back-gated devices, we show that the high $B$-field behaviour correlates with their zero *B*-field transport mobility. By this means, we extend the well-accepted notion of phonon-limited resistivity in ultra-clean graphene to a hitherto unexplored high-field realm.

**Introduction**

Van der Waals heterostructures of graphene and hBN have recently granted experimental access to novel phenomena in condensed matter [1]. The use of hBN as atomically-flat encapsulating dielectric, in particular, permits a drastic reduction of extrinsic disorder in graphene devices [2], leading to the observation of zero-field transport regimes dominated by either electron-electron [3], electron-hole [4] or electron-phonon (e-ph) interaction [5], which manifest over different carrier density and temperature ranges. Toward RT ($T$ ~ 300 K), the scattering of electrons with acoustic phonons was theoretically identified as the main intrinsic contribution to the electrical resistivity in graphene [6–8], implying a carrier mobility exceeding $10^5$ cm$^2$V$^{-1}$s$^{-1}$ at low carrier concentration ($n < 10^{12}$ cm$^{-2}$). While such figures could already be inferred from early data on disordered SiO$_2$-supported graphene (~$10^4$ cm$^2$V$^{-1}$s$^{-1}$ mobility) [9, 10], at present, the reach of the zero-field acoustic-phonon-limit is firmly established as a generic property of high-quality graphene devices [5], also when encapsulated in hBN crystals from different sources [11] or engineered to high doping levels ($n > 10^{13}$ cm$^{-2}$) [12]. Notable exceptions to the cleanness-implies-high-RT-mobility scenario are suspended graphene samples, where flexural phonons dramatically contribute to carrier scattering leading to a $T^2$ behaviour of the resistivity [13], and rotationally faulted graphene bilayers close to magic-angle, showing strong phonon-driven *T*-linear resistivity [14]. The difference between freely suspended graphene and graphene encapsulated in hBN is due to the fact that in the latter case van der Waals interaction between graphene and substrate makes flexural phonons harder, suppressing an intrinsic rippling instability [15].

In this work, we address the fundamental question whether the e-ph mechanism in clean graphene could also govern the electrical transport in the QH regime [16] at temperatures close to RT. In this sense, we note that previous literature on the RT-QH effect in graphene [17–20] exclusively includes experiments on SiO$_2$-supported devices, precluding such investigation.

**Results**

The QH effect in 2DESs manifests when the Fermi level ($E_F$) lies on the localised states between two LLs, formed in a perpendicular magnetic field and separated by an energy gap $\Delta_{LL}$. The interplay between this energy scale and the thermal energy *kT* governs the basic phenomenology of the electrical transport in the QH regime. When $kT \ll \Delta_{LL}$, no conduction takes place in the 2D bulk, while 1D chiral edge states carry the electrical current ballistically, leading to zero longitudinal resistivity ($\rho_{xx}$) when measured in four-probe configuration (Figure 1a, upper panel). As the temperature increases and $kT \sim \Delta_{LL}$, thermal excitation of extended bulk states (close to the LLs centre) exponentially restores bulk conduction and carrier scattering (Figure 1a, lower panel), resulting in a finite value of the longitudinal resistivity minimum according to $\rho_{xx} = \rho_0 \exp(-\Delta_{LL}/2kT)$. This relation is vastly employed to estimate the inter-LL separation via *T*-dependent measurements of the local resistivity minimum (under the precaution that the activation energy underestimates $\Delta_{LL}$ due to disorder-broadening of the LLs [21]). The pre-factor to the exponential term, $\rho_0$, which is often not considered explicitly, determines the magnitude of the *T*-activated resistivity (shaded yellow area Figure 1a, lower) and contains information regarding the disorder potential [22, 23]. In perpendicular magnetic fields, e-ph scattering requires lattice vibrations with a wave-vector in the order of the inverse of the magnetic length ($l_B \sim 25$ nm$/\sqrt{B[\mathrm{T}]}$) [24], which defines a third energy scale relevant to our problem $E_{ph} = \hbar v_s / l_B$ (where $v_s$ is the sound velocity in the material). In conventional 2DESs, the small $\Delta_{LL}$ leads to a complete suppression of the QH effect within a few K [25], where the $E_{ph}$-controlled phonon population can be considered negligible. Although the low electronic mass in 2DESs such as InSb [26] and HgTe [27–29] enables the observation of the QHE up to liquid-nitrogen temperature, this is insufficient to ensure $kT \gg E_{ph}$ and therefore insufficient to realize a predominance of e-ph interaction. This condition, as sketched in Figure 1b, is instead fulfilled by graphene in the RT-QH regime (the field dependence of $E_{ph}$ and the corresponding $T$-dependent excitation probability for acoustic phonons in graphene at $B$ = 30 T are shown in SI Figure S1). Under this circumstance, the $T$-activated resistivity (shaded dark cyan area in Figure 1b) should directly relate to e-ph scattering [24].

Figure 1c shows a representative measurement of the RT-QH effect, acquired at $B$ = 30 T in a hBN/graphene/hBN back-gated Hall bar (sample D2). The Hall conductivity ($\sigma_{xy}$) presents weak slope changes around filling factors $v$ = ±2 ($V_g$ ~ ±20 V), while the shelves-like features at low carrier concentration originate from the onset of electron-hole coexistence in the highly-degenerate $N$ = 0 LL [30]. $\rho_{xx}$, in addition to the pronounced maximum around the charge neutrality point (CNP), shows two sizable minima (Figure 1c, inset), indicative of $T$-activated QH states. Notably, the overall robustness of the RT-QH signatures dramatically differs in high-mobility graphene with respect to $SiO_2$-supported samples [17]; we thoroughly address this striking observation in a separate work, where we study the suppression of the $\sigma_{xy}$ plateaus in ultra-high-quality devices at temperatures significantly lower than RT. In the following, we will focus on the magnitude of $\rho_{xx}$ in the RT-QH regime and identify the underlying mechanism employing a collection of dry-assembled hBN/graphene/hBN heterostructures.

In Figure 2 we present the main transport characteristics of our devices (details on the fabrication are given in Methods), measured at zero magnetic field and at elevated temperatures. Figure 2a shows the RT mobility of three hBN-encapsulated devices, calculated according to the Drude model ($\mu = 1/(ne\rho_{xx})$), as a function of the carrier density $n$. All the mobility curves are well above the typical values for $SiO_2$-supported graphene (grey shaded area) over the whole $n$ range. Importantly, sample D3 shows a $\mu(n)$ dependence comparable to the data of Ref. [5] (dash-dotted line), demonstrating the standard fingerprint of phonon-limited RT mobility in zero magnetic field [11, 12] (as confirmed by temperature-dependent resistivity data shown in SI Figure S2). We note that, although Wang *et al.* employed a 15 µm-wide van der Pauw device, e-ph scattering imposes a ~1 µm upper bound to the electronic mean free path at $B = 0$ and RT [5]. Therefore, the zero-field e-ph limit can also be realized using narrow Hall bars, provided that their channel width exceeds 1 µm (1.5 µm to 2.3 µm in our devices). The overall high quality of the samples is also supported by the observation of fractional QH states at liquid-helium temperature (see data for sample D2 in SI Figure S3, and Ref. [31] for sample D4, fabricated using CVD-grown graphene). In Figure 2b we explore the correlation between the carrier mobility (calculated using the field-effect formula [32]) and the charge inhomogeneity in the CNP region, estimated as the usual $n^*$ parameter [33] (see Figure 2b inset for an example of the extraction). We consider data at $T$ = 220 K, where clear thermal activation is observed in the RT-QH regime. $n^*$ values above the intrinsic CNP thermal broadening (~$2.6 \times 10^{10}$ cm$^{-2}$ at 220 K, beginning of the x-axis in Figure 2b) quantify the residual disorder, which, in our devices, remains well below the typical observations for graphene on SiO$_2$ ($n^*$ in the few-$10^{11}$ cm$^{-2}$ range). In addition, as for Refs. [33, 34], the linear $\mu^{-1}(n^*)$ dependence (see shaded area in Figure 2b) indicates scattering from long-range potentials, attributed to random strain variations generic to graphene on substrates [35]. We can therefore conclude that the devices at disposal (i) span a low-disorder range unexplored in previous RT-QH experiments, and (ii) present a well-defined disorder type, with increasing impact along the D4-to-D1 sequence.

We then employ the sample temperature as an experimental knob to control the excitation of both phonons (see SI Figure S1) and bulk-extended electronic states in strong magnetic fields. In Figure 3a we sketch the effect of increasing $T$ on the Landau-quantized electrons in graphene at $B$ = 30 T. Toward RT, the broadening of the Fermi-Dirac distribution around $E_F$ (experimentally set by $V_g$) ensures excited charge carriers from both the $N$ = 0 and $N$ = 1 LLs, across the giant gap $\Delta_{LL}$. Accordingly, the local resistivity minimum at filling factor $\nu$ = 2 leaves zero and displays increasing finite values, as shown in the experimental curves of Figure 3b. In Figure 3c, we present a complete picture of the $T$-dependence of $\rho_{xx}(\nu = 2)$ for samples D1-4, at selected magnetic fields (30 T and 25 T in the main panel and inset, respectively; data at $\nu$ = -2 are shown in SI Figure S4). In addition to our data, we show reference points from Ref. [20] (black diamonds, $\rho_{xx}(\nu = 2)$ in graphene on SiO$_2$), and two theoretical calculations defining different dissipation limits (continuous lines). In both cases we take an activation energy equal to $\Delta_{LL}/2$: this was shown to be accurate for high $B$-fields in Ref. [20] and should hold true for clean graphene with reduced LL broadening. The upper line (yellow) assumes the universal conductivity pre-factor due to long-range disorder ($2e^2/h$) [23], multiplied by a factor 4 to take into account the LL degeneracy of graphene. The lower line (dark cyan) is based on the work by Alexeev *et al.* [24], who calculated the conductivity mediated by two-phonon scattering for graphene in the RT-QH regime. The relevant e-ph process conserves the LL number, but modifies the in-plane electronic momentum. We note that this phenomenology is fundamentally different from that of magneto-phonons oscillations, recently discovered in extra-wide graphene devices [36], which rely on resonant inter-LL scattering at $T < 200$ K. Here, two-phonon scattering within each LL contributes with a conductivity pre-factor $\sigma_0 = \sigma_N(T/300\text{ K})(B/10\text{ T})^{1/2}$, which depends both on temperature and magnetic field (in contrast to the constant pre-factor commonly assumed in QH studies). In the $v = 2$ state, the predominant contribution to the $\sigma_N$ terms comes from the $N = 0$ LL (0.65 $e^2/h$, one order of magnitude larger with respect to $N = 1$, 0.06 $e^2/h$) [24]. Strikingly, the resulting activated behaviour, not including any free parameter, is well approximated by our devices, while the reference data from graphene on $SiO_2$ follow the long-range disorder limit. The qualitative agreement between theoretical calculations and experimental data, together with the contrasting behaviour with respect to previous reports [20], indicate that graphene/hBN heterostructures support an e-ph-dominated transport in the RT-QH regime. Arrhenius-type fits to the conductivity [37], shown in SI Figure S5, confirm the contrasting magnitude of the pre-factor for the two generations of graphene devices (as well as the correctness of the assumed gap size).

Despite the presence of long-range potentials (Figure 2b), our data clearly indicate that the e-ph pre-factor does not simply add up to the standard long-range disorder term. To elucidate this point, we quantitatively analyse the deviation from the phonon-mediated limit in the different devices. We proceed by fitting the data from samples D1-3 (only two high $T$ curves are acquired for D4 due to experimental limitations) with a generalized relation (Figure 4, inset), which adds to the theoretical e-ph dependence from Ref. [24] an activation part with a constant pre-factor ($\rho_D$). This term is intended to account for the effect of residual disorder, and it is the only free parameter in the fits. In Figure 4 we plot the extracted $\rho_D$ for the three samples at different magnetic fields, as a function of the *n\** parameter (averaged between the electron and hole-side). The linear $\rho_D(n^*)$ behaviour observed here (shaded area in Figure 4) indicates that the random strain variations inducing the CNP broadening are also responsible for $\rho_{xx}$ exceeding the e-ph limit in the RT-QH regime. Notably, the only device to display an exact e-ph-type dependence (D3, $\rho_D \sim 0$), is also the one to show a Drude mobility comparable to the zero-field e-ph limit [5]. Taking into account the sample-dependent correction due to residual disorder, in SI (Figure S6) we proceed to a quantitative investigation of the field and temperature dependence of the conductivity pre-factor in our samples, revealing the expected $B^{1/2}$ behaviour of the e-ph term. However, we note that the simplified pre-factor proposed in Ref. [24] is the result of several approximations and, more importantly, it neglects the effect of disorder. To better understand the interplay between the different scattering mechanisms underlying the activated resistivity, in SI (Figures S7 and S8) we discuss additional data at lower temperature (down to 50 K) and magnetic field (down to 1 T). We find that $\rho_D$ drastically increases toward low *T*, with the activated resistivity exceeding the e-ph limit by more than one order of magnitude in a clean sample. However, as the temperature and magnetic field are increased, $\rho_D$ progressively drops (i. e., the activated resistivity tends toward the e-ph limit), suggesting a temperature-driven crossover between regimes dominated by either disorder or e-ph interaction (the latter being realized only close to RT). While it is not surprising that the e-ph limit works as a lower bound to the activated resistivity of real samples, the non-universality (i.e., the sample and temperature dependence) of the disorder contribution deserves particular attention in future theoretical treatments of the RT-QH in graphene.

**Discussion**

The physics of graphene is essentially determined by its deviations from flatness (that is, ripples), due to either thermal fluctuations associated to flexural phonons for freely suspended samples or to roughness of substrate like for graphene on $SiO_2$ [15]. In both cases, ripples induce inhomogeneity of electron density with electron and hole puddles in the vicinity of the CNP [38, 39]. In particular, for the case of graphene on $SiO_2$ the amplitude of induced inhomogeneity of charge-carrier density is estimated as $3\times10^{11}$ cm$^{-2}$ [39], in agreement with the above cited experimental values of *n\**. This makes the system strongly disordered, and any intrinsic scattering mechanisms become irrelevant. Oppositely, the hBN substrate is atomically flat [1] and at the same time suppresses intrinsic ripples which increases the RT carrier mobility by an order of magnitude and makes intrinsic scattering mechanisms dominant [15]. Indeed, experimentally measured *n\** for our samples is an order-of-magnitude smaller than what is supposed to be induced by ripples at RT. This results in an essentially different picture of QH physics at high enough temperatures.

In conclusion, we showed experimental evidence of predominant e-ph scattering in the QH regime. This is realized by uniquely combining strong magnetic fields, high temperatures and hBN-encapsulation of graphene. Although the RT-QH in graphene has long been known, we showed that mitigation of disorder via van der Waals engineering provides novel insights on the transport mechanisms in this phenomenon.

## Methods

*Graphene-hBN van der Waals assembly and device fabrication*

hBN/graphene/hBN samples D1-3 are assembled using the standard van der Waals dry pick-up [5], starting from micromechanically exfoliated graphene flakes previously identified by optical and Raman microscopy. Sample D4 is obtained by CVD growth on Cu foil and direct hBN-mediated pick-up after controlled decoupling via Cu surface oxidation [31]. All the devices are fabricated making use of electron beam lithography, reactive ion etching and e-beam evaporation of Cr/Au 1D edge contacts [5].

*Magnetotransport measurements*

We use standard lock-in acquisition at low frequency (13 Hz), with simultaneous $\rho_{xx}$ and $\rho_{xy}$ measurements in four-probe configuration, either under a constant current excitation (12.5 nA, sample D1-D3) or a constant voltage bias (300 µV, sample D4). The devices are mounted in a VTI system with low-pressure $^4$He serving as exchange gas, coupling the samples to a liquid-$N_2$ reservoir. The cryogenic system is accommodated in the access bore of a resistive Bitter magnet at HFML-EMFL, with a maximum field of 33 T.

## Data Availability

The data presented in this study are available at https://doi.org/10.5281/zenodo.7352031 .

**Acknowledgements**

We acknowledge technical support from Y. Lechaux and J. Quereda. This work has been supported by Ministerio de Ciencia e Innovación (Grant PID2019-106820RB-C2-2) and Junta de Castilla y León (Grants SA256P18 and SA121P20, including EU/FEDER funds). This work was supported by HFML-RU/NWO-I, member of the European Magnetic Field Laboratory (EMFL). This work was also supported by CENTERA Laboratories in the frame of the International Research Agendas Program for the Foundation for Polish Sciences co-financed by the European Union under the European Regional Development Fund (no. MAB/2018/9). D.V. acknowledges financial support from the Ministry of Universities (Spain) (Ph.D. contract FPU19/04224). J.A.D-N thanks the support from the Universidad de Salamanca for the María Zambrano postdoctoral grant funded by the Next Generation EU Funding for the Requalification of the Spanish University System 2021–23, Spanish Ministry of Universities. K.W. and T.T. acknowledge support from the Elemental Strategy Initiative conducted by the MEXT, Japan (Grant Number JPMXP0112101001) and JSPS KAKENHI (Grant Numbers 19H05790, 20H00354 and 21H05233).

This version of the article has been accepted for publication, after peer review, but is not the Version of Record and does not reflect post-acceptance improvements, or any corrections. The Version of Record is available online at: https://doi.org/10.1038/s41467-023-35986-3 .

**Author Contributions Statement**

U.Z., S.W. and S.P. conceived the experiment and coordinated the collaboration. D.V., V.C. and M.S. fabricated the graphene devices and performed the transport measurements. J.A.D.-N., A.M.-R. and J.S.-S. provided technical assistance in the cleanroom processing. C.S.A.M. and K.R. provided technical assistance during the high-field experiments. K.W. and T.T. provided single crystals of hBN. B.B., C.S. and E.D. supervised the experimental work. D.V., V.C., M.S., and S.P. performed the data analysis. M.I.K. provided theoretical input for the interpretation of the results. S.P. wrote the manuscript with input from all the co-authors.

**Competing Interests Statement**

The authors declare no competing interests.

**Figures and Captions**

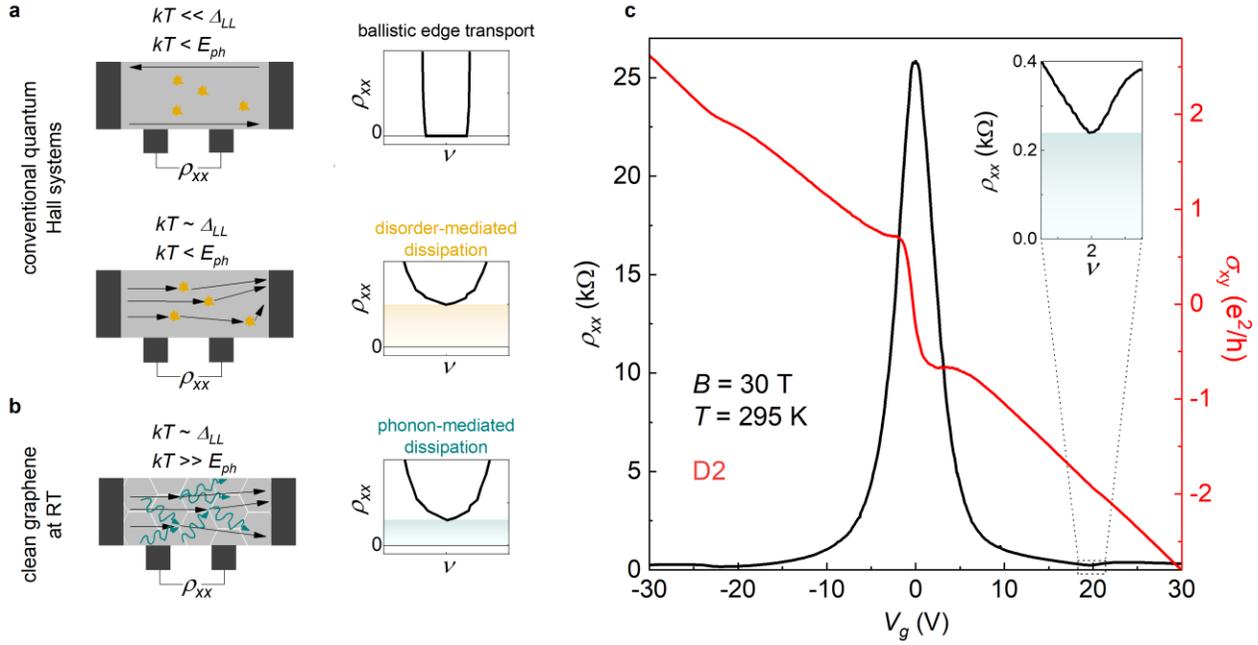

**Figure 1 | Dissipation regimes in the quantum Hall phase: high-quality graphene at RT. a,** Schematics of temperature-dependent transport in conventional quantum Hall systems, such as 2DESs in semiconductors. At low $T$ (relative to the LL separation, upper part), the electrical current is carried by chiral edge states, leading to zero longitudinal resistance. At higher $T$ (lower part), thermally-excited bulk states give a finite resistivity due to disorder scattering (yellow shading), with negligible contribution from lattice vibrations. **b,** At RT, graphene supports both the QH effect (due to large inter-LL spacing) and predominant e-ph scattering in high-mobility samples, enabling the realization of a different transport regime, with phonon-mediated dissipation at high magnetic fields (dark cyan shading). **c,** $\rho_{xx}$ (black) and $\sigma_{xy}$ (red) as a function of the back-gate voltage (corrected by a 5.2 V offset from the CNP), measured in hBN-encapsulated sample D2 at $B$ = 30 T and $T$ = 295 K. Inset: zoom-in of $\rho_{xx}$ in the vicinity of filling factor $\nu$ = 2 (the dark cyan shading indicates the finite value of the resistivity minimum).

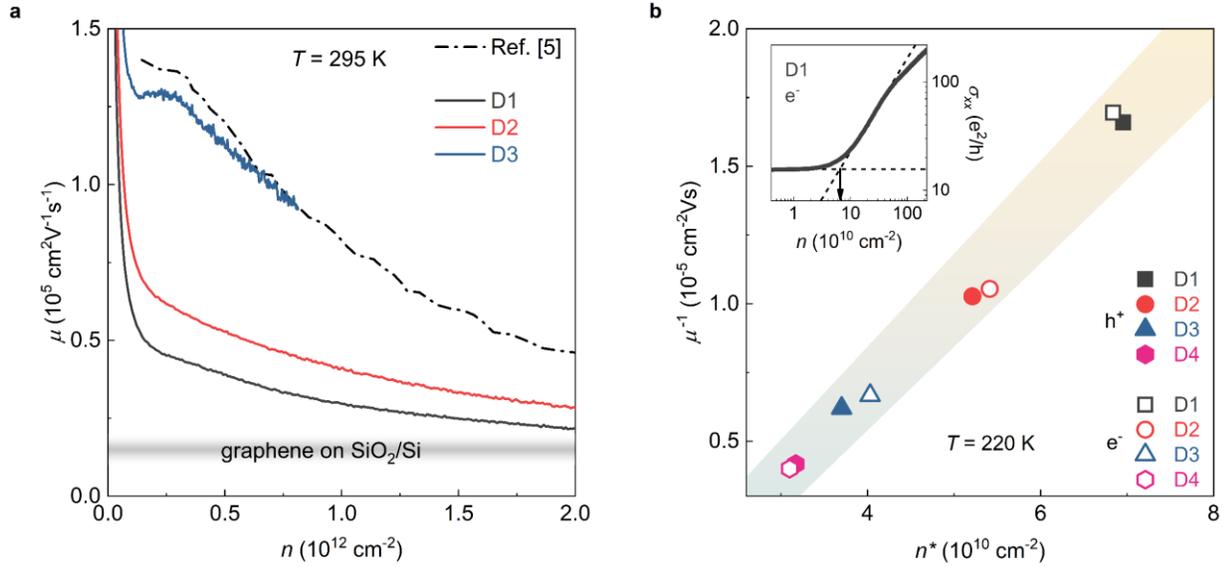

**Figure 2 | Phonon-limited transport and residual disorder at zero magnetic field. a,** RT carrier mobility (calculated according to the Drude model) as a function of the carrier concentration, for three hBN-encapsulated devices. The reference dash-dotted line are data from Ref. [5], indicating a carrier mobility limited by electron-acoustic phonon scattering. The grey-shaded area shows the typical mobility for $SiO_2$-supported graphene devices, $1\text{-}2 \times 10^4$ $cm^2V^{-1}s^{-1}$. **b,** Inverse of the high-temperature (220 K) field-effect mobility as a function of charge inhomogeneity $n^*$, for hBN/graphene/hBN devices D1-4. The shaded area covers a linear fit to the data, as in Ref. [33], ± one standard error on the best-fit intercept and slope. Inset: Log-Log plot of the longitudinal conductivity of sample D1 as a function of the carrier density, exemplifying the extraction of $n^*$ (black arrow).

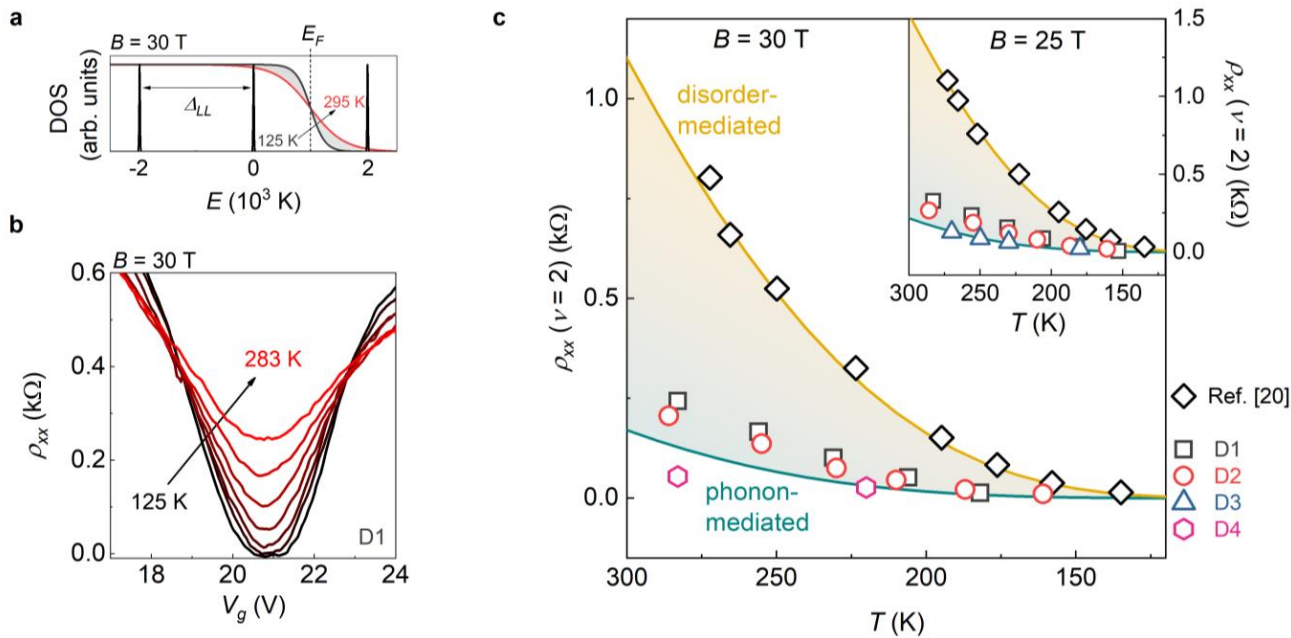

**Figure 3 | Temperature-activated resistivity and phonon-mediated dissipation in the quantum Hall effect. a,** Density of states (DOS) of graphene as a function of energy, at $B$ = 30 T (with a realistic value of LL broadening of 15 K). On top of the DOS we show the Fermi-Dirac distribution, with $E_F$ positioned in the middle of the $N$ = 0 and $N$ = 1 LL, at two different temperatures, representative of the experimental range considered. **b,** Temperature-activated longitudinal resistivity in the vicinity of $v$ = 2, measured in sample D1. **c,** Minimum of $\rho_{xx}$ at $v$ = 2 as a function of temperature, for the hBN-encapsulated devices. The reference data (black diamonds) are from Ref. [20]. The yellow and dark cyan continuous line are theoretical calculations based on Ref. [23] and Ref. [24], respectively (the shading covers resistivity values within the two theoretical calculations). The magnetic field is 30 T (25 T) in the main panel (inset).

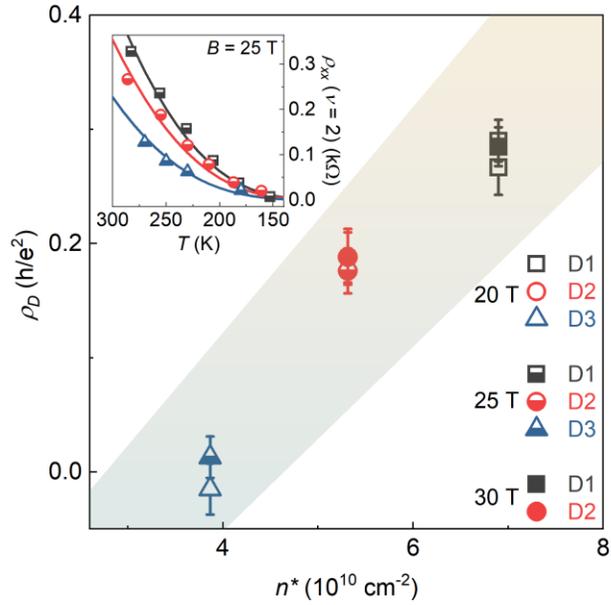

**Figure 4 | Sample-dependent disorder contribution to the activated resistivity.** Correlation between the *T*-independent pre-factor to the activated resistivity and $n^*$(220 K) for devices D1-3. The shaded area is defined as in Figure 2b. The error bars correspond to ± one standard error from the fits shown in the inset. Inset: fit to the minimum resistivity as a function of temperature (continuous lines), using the generalized formula including both e-ph and disorder contributions, at $B$ = 25 T.